\documentclass[prd,showpacs,nofootinbib,superscriptaddress,preprintnumbers]{revtex4}

\usepackage{graphicx}
\usepackage[centertags]{amsmath}
\usepackage{amsfonts}
\usepackage{amssymb}
\usepackage{amsthm}
\usepackage{newlfont}
\usepackage{subfigure}
\usepackage{multirow}
\usepackage{accents}
\usepackage{color}

\def\bea{\begin{eqnarray}}
\def\eea{\end{eqnarray}}
\def\be{\begin{equation}}
\def\ee{\end{equation}}

\usepackage{graphicx}

\def\be{\nopagebreak[3]\begin{equation}}
\def\ee{\end{equation}}
\def\ba{\nopagebreak[3]\begin{eqnarray}}
\def\ea{\end{eqnarray}}

\newcommand{\teta}{\rlap{\lower2ex\hbox{$\,\tilde{}$}}\eta{}}

\newcommand{\mr}{\mathrm}

\usepackage{enumerate}

\usepackage{colordvi}

\usepackage[centertags]{amsmath}
\usepackage{amsfonts}
\usepackage{amssymb}
\usepackage{amsthm}
\usepackage{dsfont}
\usepackage{newlfont}
\usepackage{hyperref}
\def\be{\begin{equation}}
\def\ee{\end{equation}}

\newcommand{\md}{{\mathrm d}}

\begin{document}
\preprint{\vbox{\baselineskip=12pt \rightline{CCM-18-5}
}}

\title{Hamiltonian analysis of a topological theory in the presence of boundaries}
\author{Alejandro Corichi}\email{corichi@matmor.unam.mx}
\affiliation{Centro de Ciencias Matem\'aticas, Universidad Nacional Aut\'onoma de
M\'exico, UNAM-Campus Morelia, A. Postal 61-3, Morelia, Michoac\'an 58090,
Mexico}

\author{Tatjana Vuka\v{s}inac}
\email{tatjana@umich.mx}
\affiliation{Facultad de Ingenier\'\i a Civil, Universidad Michoacana de San Nicol\'as de Hidalgo, 
Morelia, Michoac\'an 58000, Mexico}

\begin{abstract}
We perform the canonical Hamiltonian analysis of a topological gauge theory, that can be seen both as a theory defined on a four dimensional spacetime region with boundaries --the bulk theory--, or as a theory defined on the boundary of the region --the boundary theory--. In our case the bulk theory is given by the 4-dimensional $U(1)$ Pontryagin action and the boundary one is defined by the $U(1)$ Chern-Simons action.  
We analyse the conditions that need to be imposed on the bulk theory so that the total Hamiltonian, smeared constraints and generators of gauge transformations be well defined (differentiable) for generic boundary conditions. We pay special attention to the interplay between the constraints and boundary conditions in the  bulk theory on the one side, and the constraints in the boundary theory, on the other side. We illustrate how both theories are equivalent, despite the different canonical variables and constraint structure, by explicitly showing that they both have the same symmetries, degrees of freedom and observables.
\end{abstract}

\maketitle

\section{Introduction}

It is well known that gauge theories defined on spacetime regions with boundary can have degrees of freedom and observables localized on the boundary (see for example \cite{BCE}). Among them are topological theories that generate effective theories on the boundary. There is an extensive work on 3-dimensional Chern-Simons theory that generates  a Wess-Zumino-Witten theory on the boundary \cite{Witten,EMSS,MS,Banados1996,Blagojevic,BR,Troessaert,ABOL,GM,BBGS1,BBGS2}. Another example is a 5-dimensional Abelian Chern-Simons theory to which it corresponds an effective 4-dimensional theory on the boundary \cite{FPR,GS}. Yet another example is a 4-dimensional BF theory in a bounded region whose surface observables were studied in \cite{HM,Troessaert,BBTS}. 
In all of these cases the bulk theory is topological, in the sense that it does not have any local degrees of freedom, but there are degrees of freedom on the boundary. Even in the case of non-Abelian higher dimensional Chern-Simons theories (that are {\it not} topological) or  4-dimensional gravity in spacetimes with boundaries, there are induced degrees of freedom on the boundary \cite{HM,BGH}. Here we will consider the simplest case when the bulk theory and the boundary one are straightforwardly related, namely when  the Lagrangian density of the bulk theory is a total derivative. 

One well known fact about topological theories of some kind, namely when the Lagrangian density can be 
written as a total derivative --and can thus be integrated by parts into a surface integral--, is that 
they are ``equivalent" to that theory on the boundary. While there is a clear equivalence as suggested by 
the equality of their actions, one can still ask how, and in which sense are the two theories equivalent. 
Note that this is a very old question that can be put in the category of a ``classical holographic 
principle" (to distinguish it from the modern holographic principle widely studied in the literature). 
There is, of course, an ample literature on the subject (see for instance \cite{Troessaert} for a recent 
review, and references therein). 
An analysis of the covariant formulation of such topological theories has been performed before and it is 
rather easy to see that, in the covariant Hamiltonian formalism (where the phase space is defined by the 
solutions to the classical equations of motion), the theory on the bulk reduces to that on the boundary in 
a rather natural manner; the symplectic structure and therefore any Hamiltonian observable of the bulk 
theory only have contributions from the fields at the boundary (see for instance \cite{CVZ2} for an 
overview). There is no contribution from the bulk in the relevant integrals. 

The natural question that arises then is how to compare the two theories from the canonical Hamiltonian 
perspective. It is easy to see that there exists a potential mismatch even in the definition of the 
fundamental variables of the theory. While the ``configuration variables" on the boundary might be 
naturally defined by the evaluation of the corresponding bulk variables on the boundary, the same is not 
true of the canonically defined momenta. Even more, the constraints and their corresponding algebra might 
be different: Are they first/second class? Do they generate the same gauge symmetries? 
If the two theories under consideration posses these distinct features, one might then ask in which sense 
are the two theories equivalent. 
The purpose of this manuscript is to consider all these issues for the simplest of such theories, namely 
the $4D$ $U(1)$ Pontryagin action on the bulk region and the corresponding boundary theory defined by a 
$U(1)$ Chern-Simons action. 
As we shall see in detail, the canonical analysis of this theory, viewed as a bulk theory or a boundary 
one, shows that the structure and algebra of constraints are different in these two approaches. Thus it 
becomes a pressing question how to compare these two theories. To offer such comparison is the objective 
of this manuscript.

Apart from the differences we have mentioned between the two theories, there is another important issue 
that needs to be addressed. When one is considering a theory with boundaries, it is fundamental to 
consider boundary conditions for the dynamical variables. These conditions might me physically motivated, 
or can be chosen to render the theory self-consistent. In the covariant approach this entails to having a 
well defined action principle, which means the action should be differentiable (see, for instance 
\cite{CVZ2,Troessaert}). If the original action is not differentiable with the chosen boundary conditions, 
one might need to supplement it with a boundary term.
In the canonical Hamiltonian description, one adopts a slightly modified viewpoint. One starts with the 
action and through the standard Dirac procedure one obtains canonical momenta, possible primary 
constraints and a Hamiltonian. Then one imposes  that all relevant quantities be differentiable. This 
condition might require the introduction of boundary conditions, extra boundary terms, or both (see for 
instance \cite{Regge&Teitelboim}  and \cite{Troessaert} for a review). On the other hand the corresponding 
boundary theory is defined on a  region without a boundary (the boundary of the boundary is zero), so 
there are not such issues arising in its canonical formulation.

Here we shall explore the relation between the two approaches, bulk and boundary, and show that there is 
an interplay between the boundary conditions that need to be imposed in the bulk theory in order to be 
well defined in the presence of boundaries, and the constraints in the induced boundary theory. This 
result can be seen as an indication that the boundary conditions could be treated as second class 
constraints in the phase space (as explored, for instance, in \cite{Zabzine,Sheikh,Vergara}). A purely 
geometrical approach to the analysis of gauge theories in spacetimes with boundary has been developed in 
\cite{GNH,Barbero}.     

As we shall show in detail, one can see in a clear fashion that
the two theories are equivalent in a precise sense. This will involve studying in detail the physical 
degrees of freedom of the theories, through their physical observables, and the generators of interesting 
gauge symmetries such as internal gauge and spacetime diffeomorphisms. The canonical analysis of the 
Pontryagin and the Chern-Simons theories is not new, of course. For instance, it has been performed in 
\cite{EC}, and some of their results are closely related to ours. There are, however, important technical 
and conceptual differences, given that our approach to comparing both theories is different. We shall 
elaborate in sections below. The extension to a non-Abelian gauge group is straightforward and does not 
shed any new insights. For simplicity and in order to focus on the conceptual issues we shall only 
consider the Abelian theory.  

The structure of the paper is as follows. In Section~\ref{sec:2} we present the Hamiltonian analysis of 
the Pontryagin theory on a 4-dimensional spacetime region with boundary. We show that there are three 
equivalent canonical Hamiltonians and analyse the boundary conditions that one needs to impose in order to  
have a well defined total Hamiltonian. We also construct gauge generators and a set of observables for 
this theory. In Section~\ref{sec:3} we perform the canonical decomposition of the Chern-Simons theory 
defined on the 3-dimensional ``time-like" boundary, and also construct the corresponding gauge generators 
and a set of observables. In Section.~\ref{sec:4} we summarize the results and show the relation between 
canonical descriptions of the two theories. An outlook is the subject of Sec.~\ref{sec:5}.

In this manuscript we adopt the {\it abstract index notation} of Penrose, where geometrical objects are 
equipped with an abstract index that does not ``take any values". Thus, while for instance a connection 
$A_a$ on spacetime and
its pullback to the boundary will be denoted by the same symbol $A_a$, it should be clear, we hope, that 
the two objects are defined on different manifolds. We shall not use a new symbol to denote pullbacks or 
restrictions for notational simplicity, and hope the meaning shall become clear from the context.
We shall use a fully covariant approach in the canonical decomposition, without fixing a 
foliation nor a coordinate system. We have kept the mathematical sophistication to the level of the 
theoretical physics literature, without the
rigour of mathematical physics  employed, for instance, in \cite{Barbero}. 

\section{Hamiltonian analysis of the Pontryagin theory}
\label{sec:2}

In this section we perform the canonical Hamiltonian analysis of the theory in the bulk, namely 
of the Pontryagin theory on the four dimensional spacetime region ${\mathcal{M}}$, with boundary, which we shall take to be of the form ${\mathcal{M}}=I \times \Sigma$, with $I$ a closed interval. $\Sigma$ is a three dimensional manifold with boundary $\partial\Sigma$ that we shall take to have the topology of a two sphere $S^2$. This section has three parts. In the first one, we perform the canonical 3+1 decomposition of the theory, following the Dirac algorithm. In the second part we construct physical observables out of simple expressions and in the third part we construct the generators of gauge symmetries.

\subsection{Canonical decomposition}

The Pontryagin action for the Abelian theory is
\begin{equation}\label{Pontryagin}
S_{\mr{P}} =   \int_{\mathcal{M}} {\mathbf{F}}\wedge {\mathbf{F}} =  \frac{1}{4} \int_{\mathcal{M}} \md^{4} x\, \tilde{\varepsilon}^{abcd}F_{ab} F_{cd}\, ,
\end{equation}
where ${\mathbf{F}}=\md {\mathbf{A}}$ is the field strength two-form, the curvature of the $U(1)$ connection one-form ${\mathbf{A}}$ and $\tilde{\varepsilon}^{abcd}$ is the Levi-Civita tensor density.
Here we have set the dimension-full parameter $\theta=1$, since there is no loss of generality as we are not coupling the fundamental variables to another field. 
We introduce a foliation of the spacetime $\mathcal{M}=I\times\Sigma$  and a ``time'' 
function $t$ such that hypersurfaces $\Sigma_t$, that correspond to $t = {\mathrm{const}}$, are diffeomorphic to $\Sigma$. Then we can introduce a time evolution vector field 
$t^{a}:=(\frac{\partial}{\partial t})^a$ such that $t^{a} \partial_{a} t = 1$. Using the identity   $\tilde{\varepsilon}^{abcd} = 4t^{[a} \tilde{\varepsilon}^{ bcd]} $ 
(that implies $\tilde{\varepsilon}^{abcd}t_a=\tilde{\varepsilon}^{bcd}$, where $t_a:=\partial_a t$), the action can be expressed as
\begin{equation}
S_{\mr{P}} = \int_{\mathcal{M}} \md^4 x\,  \tilde{\varepsilon}^{abc} (t^d F_{da})\, F_{bc} 
=  \int_{\mathcal{M}}\md^4 x\,  \tilde{\varepsilon}^{abc} F_{bc} \left( \pounds_{t} A_{a} - \partial_{a} \phi\right)\, ,
\end{equation}
where $\phi :=t\cdot \mathbf{A}$ and we have used the Cartan identity   $\pounds_{t} A_{a} =t^{b} F_{ba} + \partial_{a} 
( t \cdot \mathbf{A} )$.
Then, the canonical momenta is given by
\begin{equation}
\tilde{\Pi}^{a} := \frac{\delta \mathcal{L}}{\delta (\pounds_{t} A_{a})} =  \tilde{\varepsilon}^{abc} F_{bc} \,  ,
\end{equation}
where $\mathcal{L}$ is the Lagrangian density.
It should be noted that the momenta are intrinsic to $\Sigma$, since $t_a\tilde{\Pi}^{a}=0$. 
The kinematical phase space is 8-dimensional (per point) and is parametrized by
the set of canonical variables  $(\phi ,\tilde{\Pi}_{\phi}; A_a,\tilde{\Pi}^a)$ where $\tilde{\Pi}_{\phi}:=t\cdot\tilde{\mathbf{\Pi}}$ and $A_a$ is the pullback of 
${\mathbf{A}}$ to $\Sigma$.

The theory has four primary constraints
\begin{eqnarray}
\tilde{\mathcal{C}} &:=& \tilde{\Pi}_{\phi}\approx 0\, ,\\
\tilde{\mathcal{C}}^a &:=& \tilde{\Pi}^a- \tilde{\varepsilon}^{abc} F_{bc}\approx 0\, .
\end{eqnarray}
Note that $t_a\mathcal{C}^a=0$, so  $\mathcal{C}^a$ is intrinsic to $\Sigma$.

The canonical Hamiltonian is defined through the Legendre transformation,
\begin{equation}\label{CanonicalHamiltonianGeneral}
H_{\mr{C}} = \int_{\Sigma} \md^{3}x\,  \left[ (\pounds_t\phi )\tilde{\Pi}_{\phi} + 
(\pounds_{t} A_{a} ) \tilde{\Pi}^{a} - \mathcal{L}  \right]
= \int_{\Sigma} \md^{3}x\,  \left[ (\pounds_t\phi )\tilde{\Pi}_{\phi}+ (\pounds_{t} A_{a} ) \tilde{\Pi}^{a}- 
\tilde{\varepsilon}^{abc} F_{bc} \left( \pounds_{t} A_{a} - \partial_{a} \phi\right)\right]\, .
\end{equation}
Since the Lagrangian is linear in $\pounds_{t} A_{a}$, this velocity term cannot be expressed as a function of $\tilde{\Pi}^{a}$. As a result we can write the canonical Hamiltonian in several
different forms that we are going to explore in detail in the following sections. 

The general strategy for the canonical analysis of the constrained system will be the following. We are going to assume that the Lagrangian theory is well defined, without paying attention to the boundary conditions that might have appeared there. Instead, we shall proceed with the canonical theory anew: we start with generic boundary conditions, and let the theory ``tell" us what modifications, in terms of boundary conditions or extra terms, might be needed. The first such conditions might come from the first step, namely in having a well defined, differentiable, canonical Hamiltonian. That is the subject of the sections that follow.   

Let us end this part with a comment. In some part of the literature, one can find that certain boundary conditions are imposed in the covariant action principle, in order to make it differentiable. These conditions are then ``carried over" to the canonical analysis and could yield, consequently, a different sector of the theory. Our viewpoint here is  to impose any consistency conditions only at the canonical level, and consider the most general variations that are allowed.

\subsubsection{First approach}

Let us start by considering the first form that the canonical Hamiltonian can take.
If we substitute $\tilde{\varepsilon}^{abc} F_{bc}=\tilde{\Pi}^a$ and $\tilde{\Pi}_{\phi}=0$ in (\ref{CanonicalHamiltonianGeneral}),  
the canonical Hamiltonian becomes
\begin{equation}\label{CanonicalHamiltonian-PontryaginV1}
H_{\mr{C1}} = \int_{\Sigma} \md^{3}x\, \tilde{\Pi}^a\partial_a\phi \, .
\end{equation}
The variation of $H_{\mr{C1}}$ on the phase space is then
\begin{equation}
\delta H_{\mr{C1}}= \int_{\Sigma} \md^{3}x\, \bigl[ (\partial_a\phi )\delta \tilde{\Pi}^a - (\nabla_a\tilde{\Pi}^a)\delta\phi\bigr]
+\int_{\Sigma} \md^{3}x\, \nabla_a (\tilde{\Pi}^a\delta\phi) \, .
\end{equation}
Note that since $\tilde{\Pi}^a$ is a vector density of weight 1, $\nabla_a\tilde{\Pi}^a=\partial_a\tilde{\Pi}^a$\footnote{From now on we shall use the symbol $\nabla_a$ for covariant derivatives, but it should be clear that all expression should be independent of the choice of
derivative $\nabla_a$, since the theory does not depend on any background structure like a metric or connection.}.
Now, we can use Stokes' theorem for vector densities of weight 1, so that
\begin{equation}
\int_{\Sigma} \md^{3}x\, \nabla_a (\tilde{\Pi}^a\delta\phi)= \int_{\partial\Sigma} \md S_a\, \tilde{\Pi}^a\delta\phi\, ,
\end{equation}
where $\md S_a =\frac{1}{2}\tilde{\varepsilon}_{abc}\md x^b\wedge \md x^c$. 
As a result, $H_{\mr{C1}}$ is a differentiable function if this boundary term vanishes for arbitrary variations $\delta\phi$. 
Let us introduce the coordinate $r$ such that $\partial\Sigma$ is defined as the surface $r={\mathrm{const}}$. and $r_a=\partial_a r$ is its normal 1-form. Then, $ \md S_a\sim r_a$ and the condition for differentiability takes the form 
\begin{equation}
r_a \tilde{\Pi}^a\vert_{\partial\Sigma}\approx  r_a\tilde{\varepsilon}^{abc}F_{bc}\vert_{\partial\Sigma} =  \tilde{\varepsilon}^{bc}F_{bc}\vert_{\partial\Sigma}=0\, .\label{cond1}
\end{equation}
This implies that the pullback of the curvature to the boundary vanishes: $F_{ab}\vert_{\partial\Sigma}=0$. 
We shall demand that these conditions hold on the whole boundary $I\times\partial\Sigma$. 
As we stated earlier, we are assuming that
the boundary of $\Sigma$, has the topology of a two sphere, $\partial \Sigma\approx S^2$. 
This in turn implies that $H^1(\partial\Sigma)=0$, so the only allowed variations of the pullback 
of $\mathbf{A}$ to the boundary are of the form, $\delta A_a = \partial_a\lambda\vert_{\partial\Sigma}$, 
where $\lambda$ is an arbitrary function on $\partial\Sigma$. Had we allowed for more complicated topologies (like a two torus) or punctures, we would have had more possibilities (and boundary degrees of freedom). 

It should be noted that in requiring that the canonical Hamiltonian be differentiable, one has to impose
boundary conditions on the connection, restricting to those that are flat on $\partial\Sigma$.

The total Hamiltonian is:
\begin{equation}\label{TotalHamiltonian-Pontryagin}
H_{\mr{T1}} = \int_{\Sigma} \md^{3}x\, (\tilde{\Pi}^a\partial_a\phi + u\, \tilde{\mathcal{C}} + u_a\,  \tilde{\mathcal{C}}^a) \, ,
\end{equation}
and it  is a differentiable function on the phase space if (\ref{cond1}) is satisfied and:
\begin{equation}
\int_{\partial\Sigma} \md^2y\, \tilde{\varepsilon}^{ab}u_a\delta A_b = 0\, .\label{cond2}
\end{equation}
This condition is  satisfied for the allowed $\delta A_a$ on the boundary if the Lagrange multipliers $u_a$ vanish at the boundary, 
$u_a\vert_{\partial\Sigma}=0$ or in the special case when $u_a=\partial_a f\vert_{\partial\Sigma}$, where $f$ is an arbitrary function on the boundary. In this second case the condition (\ref{cond2})
reduces to
\begin{equation}
\int_{\partial\Sigma} \md^2y\, \tilde{\varepsilon}^{ab}\partial_a f\partial_b\lambda=
-\int_{\partial\Sigma} \md^2y\, \tilde{\varepsilon}^{ab}f\nabla_a\nabla_b \lambda = 0\, .
\end{equation}

If either of these conditions are imposed, then the variation of the total Hamiltonian takes the form,
\begin{equation}
\delta H_{\mr{T1}}=\int_{\Sigma}\md^3 x\, \bigl[ -(\nabla_a\tilde{\Pi}^a)\delta\phi + u\, \delta\tilde{\Pi}_{\phi} 
-2\,\tilde{\varepsilon}^{abc}(\nabla_b u_c)\delta A_a + (u_a+\partial_a\phi)\delta\tilde{\Pi}^a\bigr]\, .  
\end{equation}

Let us now consider the differentiability of the smeared constraints. The first one
\begin{equation}
C_1[v] = \int_{\Sigma} \md^3 x\, v\, \tilde{\Pi}_{\phi}\, ,
\end{equation}
is clearly differentiable, since it does not contain any derivative, while the second smeared constrained
\begin{equation}
C_2[v_a] = \int_{\Sigma} \md^3 x\, v_a\, ( \tilde{\Pi}^a - \tilde{\varepsilon}^{abc}F_{bc})\, ,
\end{equation}
is differentiable only if a condition similar to (\ref{cond2}) is imposed, namely
\begin{equation}
\int_{\partial\Sigma} \md^2y\, \tilde{\varepsilon}^{ab}v_a\delta A_b = 0\, ,
\end{equation}
that again leads to $v_a\vert_{\partial\Sigma}=0$ or $v_a=\partial_a g\vert_{\partial\Sigma}$. 
Allowed test functions are the ones that satisfy one of these restrictions.  Then,
\begin{equation}
\delta C_2 = \int_{\Sigma}\md^3 x\, \bigl[ -2\tilde{\varepsilon}^{abc}(\nabla_b v_c)\delta A_a + 
v_a\delta\tilde{\Pi}^a\bigr]\, .
\end{equation}

We can now analyze the consistency conditions of the constraints by computing
\begin{eqnarray}
\{ C_1[v],H_{\mr{T1}}\}\approx 0\ \ \ &\Rightarrow& \ \ \ \tilde{\mathcal{C}}_3:=\nabla_a\tilde{\Pi}^a\approx 0\, ,\\
\{ C_2[v_a],H_{\mr{T1}}\}\approx 0\ \ \ &\Rightarrow& \ \ \ \int_{\partial\Sigma}\md^2 y\, \tilde{\varepsilon}^{ab}
v_a (u_b+\partial_b\phi ) =0\, .\label{cond3}
\end{eqnarray}
The secondary constraint $\tilde{\mathcal{C}}_3$ is not independent, since $\tilde{\mathcal{C}}_3=\nabla_a\tilde{\mathcal{C}}^a$.
The condition (\ref{cond3}) is satisfied for all allowed $u_a$ and $v_a$.
With this, we can conclude that there are no extra (independent) constraints for the system.

In order to further classify the constraints, we compute their Poisson brackets to obtain the algebra of constraints,
\begin{eqnarray}
\{ C_1[v],C_1[w]\} &=& \{ C_1[v],C_2[w_a]\}=0\, ,\\
\{ C_2[v_a],C_2[w_b]\} &=& 2\int_{\partial\Sigma}\md^2 y\, \tilde{\varepsilon}^{ab}v_a w_b = 0\, .
\end{eqnarray}
Thus, the constraints are first class for all allowed test functions. Let us now perform a quick counting of (local) degrees of freedom. We started with 8 phase space degrees of freedom $(\phi,\tilde\Pi_\phi; A_a,\tilde{\Pi}^a)$, and we have seen that there are four independent first class constraints $({\tilde{\cal{C}}},
{\tilde{\cal{C}}}^a)$. Thus, there are $8- 2\cdot 4 =0$ local degrees of freedom, which means that if there are any degrees of freedom, they would have to arise as would-be gauge degrees of freedom at the boundary.
That will, of course, depend on both the boundary conditions and on the
topology of the spacetime region. In our case we have no boundary degrees of freedom. 

Let us now repeat the Dirac analysis for a second possible version of the canonical Hamiltonian.

\subsubsection{Second approach}

We can obtain this second way of representing the Hamiltonian
if we substitute $\tilde{\Pi}^a= \tilde{\varepsilon}^{abc} F_{bc}$ and $\tilde{\Pi}_\phi  =0$ in (\ref{CanonicalHamiltonianGeneral}).   
The canonical Hamiltonian becomes
\begin{equation}\label{CanonicalHamiltonian-PontryaginV2}
H_{\mr{C2}} = \int_{\Sigma} \md^{3}x\, \tilde{\varepsilon}^{abc}F_{bc}\,\partial_a\phi 
= H_{\mr{C1}} + C_2[\partial_a\phi ]     \, .
\end{equation}
Its variation is
\begin{equation}\label{var2}
\delta H_{\mr{C2}}= -\int_{\Sigma}\md^3 x\, \tilde{\varepsilon}^{abc} \bigl[ 2(\nabla_b\nabla_a\phi )\delta A_c +
\nabla_a F_{bc}\, \delta\phi\bigr]
+\int_{\partial\Sigma} \md^{2}y\, \tilde{\varepsilon}^{ab} (-2\,\partial_a\phi\,\delta A_b + F_{ab}\,\delta\phi)\, ,
\end{equation}
so that the bulk term vanishes identically and the variation reduces to a boundary term.

The variation of $H_{\mr{T2}}$ is well defined if the boundary term in $\delta H_{\mr{C2}}$ vanishes, so we need to impose the condition (\ref{cond1}),
in this case 
\begin{equation}
\delta H_{\mr{T2}} = \int_{\Sigma}\md^3 x\, \bigl[ u\, \delta\tilde{\Pi}_{\phi} 
+2\tilde{\varepsilon}^{abc}(\nabla_b u_c)\delta A_a + u_a\delta\tilde{\Pi}^a\bigr]\, ,
\end{equation}
for all allowed multipliers. Now, the consistency conditions and algebra of the constraints can be found straightforwardly,
\begin{eqnarray}
\{ C_1[v],H_{\mr{T2}}\} &=& 0\, ,\\
\{ C_2[v_a],H_{\mr{T2}}\} &=& 0\, ,\\
\{ C_2[v_a],C_2[u_a]\} &=& 0\, .
\end{eqnarray}
Note that in this case the consistency condition of $C_1$ is identically fulfilled, it does not produce any secondary
constraint, as was the case in the first approach (even when that secondary constraint was not independent). 

\subsubsection{Third approach}

In this third approach, we perform an integration by parts in (\ref{CanonicalHamiltonian-PontryaginV2}) and obtain a contribution only from the boundary
\begin{equation}\label{CanonicalHamiltonian-PontryaginV3}
H_{\mr{C3}} = \int_{\partial\Sigma} \md^{2}y\, \tilde{\varepsilon}^{ab}F_{ab}\,\phi\, .
\end{equation}
Its variation is given by 
\begin{equation}\label{var3}
\delta H_{\mr{C3}}= 
\int_{\partial\Sigma} \md^{2}y\, \tilde{\varepsilon}^{ab} (-2\,\partial_a\phi\,\delta A_b + F_{ab}\,
\delta\phi)\, .
\end{equation}
This is the same as in (\ref{var2}), so that $H_{\mr{T3}}$ is well defined if the condition (\ref{cond1}) is satisfied and for all allowed multipliers, just like in our second approach. 

Note that the form of the canonical Hamiltonian is the same as in the Chern-Simons theory (see below), defined on $\partial\Sigma$, but since
the total Hamiltonian of the Pontryagin theory is defined on $\Sigma$ the boundary terms in its variation should vanish. 
If we had started from the Chern-Simons theory on a three dimensional manifold $I\times\partial\Sigma$,
we would have arrived at the canonical Hamiltonian (\ref{CanonicalHamiltonian-PontryaginV3}), but the momenta and the constraints
would have been different than in Pontryagin theory, and the corresponding total Hamiltonian would be defined on $\partial\Sigma$ (see below). This third approach makes it easier to compare to the case we shall consider in the next section, namely an Abelian Chern-Simons theory on $I\times\partial\Sigma$.

\subsection{Observables}

One of the main goals of this article is to compare two apparently distinct theories, namely Pontryagin on the bulk and Chern-Simons on the boundary. While we know that they are equivalent at the level of the action, we have two very different canonical descriptions, for the simple fact that they are defined on different spaces, one being the boundary of the other. One expects that, while the basic variables and the structure of the constraints might be different, one should still be able to describe the same ``physics". It is natural then to consider physical observables, and the algebra they satisfy, as a way of matching both theories. Of course this strategy is not new and is pursued in the context of, for instance, dualities {\it a la} AdS/CFT \cite{CHD,Maldacena,Witten2}. 

General considerations on diffeomorphism invariant theories allow one to conclude that physical observables, when written as integrals over a hyper-surface $\Sigma$ of an integrand that depends (locally) on the fields and finite derivatives, have contributions only from the boundaries of  
$\Sigma$ (see for instance the discussion in \cite{CVZ2}). Thus, it is natural to follow that strategy first proposed by Regge and Teitelboim \cite{Regge&Teitelboim}, as we now describe.
We shall try to construct observables as boundary terms that should be added to smeared first class constraints in order
to make them differentiable, without imposing any conditions on the multipliers. In this case, since
\begin{equation}
\delta C_2[u_a] = \int_{\Sigma}\md^3 x\, \bigl[ -2\tilde{\varepsilon}^{abc}(\nabla_b v_c)\delta A_a + 
v_a\delta\tilde{\Pi}^a\bigr]+\int_{\partial\Sigma} \md^2y\, \tilde{\varepsilon}^{ab}v_a\delta A_b\, ,
\end{equation}
we see that the functional
\begin{equation}
Q[w_a]:= C_2[w_a]- \int_{\partial\Sigma} \md^2y\, \tilde{\varepsilon}^{ab}w_a A_b\, ,
\end{equation}
is differentiable for an arbitrary one-form $w_a$. In order to be an observable its Poisson brackets with the constraints should vanish. First note that $\{ Q[w_a], C_1[v]\}=0$ and 
\begin{equation}
\{ Q[w_a], C_2[u_b]\} = 2\int_{\partial\Sigma}\md^2 y\, \tilde{\varepsilon}^{ab}w_a u_b\, .
\end{equation}
This Poisson bracket vanishes for: 1) arbitrary $w_a$ on $\partial\Sigma$ in the case when $u_b=0\vert_{\partial\Sigma}$ or 
2) for $w_a=0\vert_{\partial\Sigma}$ or $w_a=\partial_a f\vert_{\partial\Sigma}$ when $u_a=\partial_a g\vert_{\partial\Sigma}$.
As a result $Q[w_a]$ is an observable only when $w_a$ satisfy the same conditions as $u_b$ on $\partial\Sigma$.
The first case, when $w_a=0\vert_{\partial\Sigma}$, is trivial since $Q[w_a]\approx 0$. In the second case, when 
$w_a=\partial_a f\vert_{\partial\Sigma}$, we have
\begin{equation}
Q[f]\approx - \int_{\partial\Sigma} \md^2y\, \tilde{\varepsilon}^{ab}\partial_a f A_b =
\frac{1}{2}\int_{\partial\Sigma} \md^2y\, f \tilde{\varepsilon}^{ab}F_{ab}=0\, ,
\end{equation}
due to the condition (\ref{cond1}). As a result, this construction leads to trivial observables.\\

Another possible strategy for finding non trivial observables is to consider a family of linear functionals as
\begin{equation}
\pi[g_a]=\int_{\Sigma}\md^3 x\, g_a\tilde{\Pi}^a\, , 
\end{equation}
where $g_a$ is an arbitrary one-form on $\Sigma$. In this case we have
\begin{equation}
\{ \pi[g_a], C_2[v_c]\} = -2 \int_{\Sigma}\md^3 x\, \tilde{\varepsilon}^{abc}(\nabla_b v_c) g_a\, ,
\end{equation}
for all allowed $v_c$. This can be rewritten as
\begin{equation}
\{ \pi[g_a], C_2[v_c]\} = 2 \int_{\Sigma}\md^3 x\, \tilde{\varepsilon}^{abc}(\nabla_b g_a) v_c - 
2\int_{\partial\Sigma} \md^{2}y\, \tilde{\varepsilon}^{ab} v_a g_b   \, ,
\end{equation}
and this expression does not vanish unless $g_a=\nabla_a f$, that leads to the family of linear observables of the form
\begin{equation}
\mathcal{O}[f]=\int_{\Sigma}\md^3 x\, (\nabla_a f)\tilde{\Pi}^a
\approx \int_{\partial\Sigma} \md S_a\,f \tilde{\Pi}^a =0\, , 
\end{equation}
due to (\ref{cond1}), again resulting in trivial observables. \\


The other family of linear functionals could be constructed as
\begin{equation}
q[h^a]= \int_{\Sigma}\md^3 x\, h^a A_a\, , 
\end{equation}
but in this case
\begin{equation}
\{ q[h^a], C_2[v_b]\}=  \int_{\Sigma}\md^3 x\, h^a v_a\, ,
\end{equation}
so $q[h^a]$ is not an observable for any election of $h^a$.

Let us also consider a non-linear functional of the form
\begin{equation}
N[f]= \int_{\Sigma}\md^3 x\, f(\nabla_a \phi)\tilde{\Pi}^a\, .
\end{equation}
This expression is differentiable
\begin{equation}
\delta N= \int_{\Sigma}\md^3 x\, [f(\nabla_a\phi )\delta\tilde{\Pi}^a - \nabla_a(f\tilde{\Pi}^a)\delta\phi ] +
\int_{\partial\Sigma} \md S_a\, f\tilde{\Pi}^a\delta\phi\, ,
\end{equation}
since the boundary term vanishes due to (\ref{cond1}). Now,
\begin{equation}
\{ N[f], C_1[g]\} \approx - \int_{\Sigma}\md^3 x\, g \tilde{\Pi}^a \nabla_a f\, ,
\end{equation}
and it vanishes only if $f={\mathrm{const.}}$ In this case,
\begin{equation}
\{ N[f], C_2[u_b]\} \approx  2 f\int_{\partial\Sigma} \md^{2}y\, \tilde{\varepsilon}^{ab}(\nabla_a\phi )u_b\, ,
\end{equation}
and it vanishes for all allowed $u_b$. For any constant $f$, we again have a trivial observable, since $N[f]=f H_{\mr{C1}}$, and
\begin{equation}
N[f]=f  \int_{\Sigma}\md^3 x\, (\nabla_a \phi)\tilde{\Pi}^a\approx f\int_{\partial\Sigma} \md S_a\, \tilde{\Pi}^a \phi =0\, .
\end{equation}
Note that the fact $H_{\mr{C1}}\approx 0$ indicates that the theory is invariant under diffeomorphisms.

Let us examine yet another non-linear functional
\begin{equation}
P[g^a] =  \int_{\Sigma}\md^3 x\, g^a\tilde{\Pi}^b F_{ab}\, .
\end{equation}
This functional is differentiable for $r_b g^b=0\vert_{\partial\Sigma}$. Then,
\begin{equation}
\{  P[g^a], C_2[u_b]\} \approx \int_{\Sigma}\md^3 x\, g^d \tilde{\varepsilon}^{abc}
[2F_{ad}\nabla_bu_c - F_{bc} (\nabla_a u_d-\nabla_d u_a)]\, ,
\end{equation}
and it does not vanish for any $g^a \ne 0$, so it is not a physical observable.

We started with a theory that is known not to have any local degrees of freedom, so the only possibility, when there is a boundary present, is that new would-be gauge degrees of freedom 
arise on the boundary. In some cases, these have been referred to as ``edge states". By considering
a time-like boundary that has the topology $I\times S^2$, one expects a theory of Abelian flat connections on the boundary to have no degrees of freedom. That is indeed corroborated by the fact that
a small sample of simple  observables are trivial. While we have not
provided an exhaustive list of candidates, one should expect that (the gradients of) linear and the simplest non-linear observables should span, locally, the co-tangent space of the phase space. Thus, this suggests that  there are no non-trivial observables, and therefore, no local physical degrees of freedom at the boundary. As mentioned before, if the topology of the boundary were non-trivial, in the sense that $H^1(\partial\Sigma)\neq 0$, we would have non-trivial global degrees of freedom that could be explored, for instance, by Wilson-loops around homotopically non-trivial curves.


\subsection{Generator of gauge transformations}

Let us now find the generators of gauge transformations, which are
 constructed as a linear combination of first class constraints
\begin{equation}
G[\epsilon_1 ,\epsilon_2 ,\eta_a]=\int_{\Sigma}\md^3x\, \bigl[ \epsilon_1\tilde{\Pi}_\phi + 
\epsilon_2\nabla_a\tilde{\Pi}^a + \eta_a (\tilde{\Pi}^a - \tilde{\varepsilon}^{abc}F_{bc})\bigr]\, ,
\end{equation}
Though the constraint $\nabla_a\tilde{\Pi}^a\approx 0$ is not an independent one, we include it as a part of the generator, and we shall show that this proposal generates the usual gauge and diffeomorphism symmetries of the theory.  
The generator is differentiable if the condition (\ref{cond1}) is satisfied and $\eta_a\vert_{\partial\Sigma}=0$ or 
$\eta_a=\partial_a g\vert_{\partial\Sigma}$. The corresponding gauge transformations are
\begin{eqnarray}
\delta\phi &=&  \epsilon_1\, , \ \ \ \delta\tilde{\Pi}_{\phi}=0\, ,\\ 
\delta A_a &=& -\nabla_a\epsilon_2 + \eta_a\, , \\
\delta\tilde{\Pi}^a &=& 2\tilde{\varepsilon}^{abc}\nabla_b \eta_c\, .
\end{eqnarray}
Depending on the particular choices of the smearing functions, that can be phase space independent or
dependent (live), one has different classes of gauge transformations.

We shall consider three cases:
\begin{enumerate}
\item $U(1)$ gauge symmetry is obtained for the choice  $\epsilon_1= \pounds_{t}\epsilon$, $\epsilon_2=-\epsilon$ and $\eta_a=0$,
then
\begin{eqnarray}
 \delta_{\epsilon}\phi &=& \pounds_{t}\epsilon\, ,\nonumber\\
 \delta_{\epsilon} A_a &=& \nabla_a\epsilon\, .
\end{eqnarray}
Here we have used the result by \cite{Castellani:1981} that states that, in order to obtain (spacetime) gauge transformations,
one has to choose the parameter $\epsilon_1$ to be the time derivative of the parameter involved in spatial transformations.

\item Spatial diffeomorphisms are obtained for the phase space dependent choice $\epsilon_1=\pounds_{\xi}\phi$, $\epsilon_2=-\xi^a A_a$ and
$\eta_a=\xi^b F_{ba}$ (note that $\eta_a\vert_{\partial\Sigma}=0$), where $\xi^a t_a=0$.  Then, we have
\begin{eqnarray}
\delta_{\xi}\phi &=&  \pounds_{\xi}\phi\, ,\nonumber\\
\delta_{\xi} A_a &=& \partial_a(\xi^b A_b)+ \xi^b F_{ba}= (\nabla_a \xi^b) A_b+ \xi^b\nabla_b A_a =\pounds_{\xi} A_a\, .
\end{eqnarray}
On the other hand
\begin{eqnarray}
\delta_{\xi}\tilde{\Pi}^a &=& 2\tilde{\varepsilon}^{abc}\nabla_b(\xi^d F_{dc})\approx 
\tilde{\varepsilon}^{abc}\underaccent{\tilde}{\varepsilon}_{dcn}\nabla_b ( \xi^d\tilde{\Pi}^n)\nonumber\\
&=&\nabla_b(\xi^b\tilde{\Pi}^a-\xi^a\tilde{\Pi}^b)=
\nabla_b(\xi^b\tilde{\Pi}^a)-(\nabla_b\xi^a)\tilde{\Pi}^b - \xi^a\nabla_b\tilde{\Pi}^b\approx\pounds_{\xi} \tilde{\Pi}^a\, ,
\end{eqnarray}
since $\underaccent{\tilde}{\varepsilon}_{dcn}\tilde{\Pi}^n\approx 2 F_{dc}$. 
Thus, $A_a$ transforms like a one form and $\tilde{\Pi}^a$ like
a vector density of weight 1 under spatial diffeomorphisms.

\item `Time-like' diffeomorphisms, or `time evolution' are obtained for $\epsilon_1=\pounds_{t}\phi$, $\epsilon_2=-\phi$ and
$\eta_a=\pounds_{t}A_a-\nabla_a\phi$. Then, we have
\begin{eqnarray}
\delta_t\phi =  \pounds_{t}\phi\, ,\nonumber\\
\delta_t A_a =\pounds_{t} A_a\, .
\end{eqnarray}
It is straightforward to check that the transformations for the canonical momenta have similar expressions.

\end{enumerate}

In the next section we shall consider the theory as a boundary theory defined on $I\times\partial\Sigma$.

\section{Hamiltonian analysis of the Chern-Simons theory on the boundary}
\label{sec:3}

Integration by parts in the Pontryagin action (\ref{Pontryagin}), leads to the action of
the Chern-Simons theory on the boundary $\mathcal{B}= I\times\partial\Sigma$ given by 
\begin{equation}
S_{\mathrm{CS}} = - \frac{1}{2}\int_{\mathcal{B}}\md^3 x\, \tilde{\varepsilon}^{abc} A_a F_{bc}\, .
\end{equation}
Even when the boundary of the spacetime region where the Pontryagin theory is defined includes the initial and final ``spatial" hypersurfaces, it is customary to neglect the contributions to the action from those hypersurfaces since in the action principle the variations of the fields are always vanishing. We shall take this viewpoint here.
This section has three parts. In the first one, we perform the canonical decomposition of the Chern-Simon theory on the ``time-like" boundary.\footnote{Note that we are writing ``space-like" and ``time-like" in analogy with theories where a metric exists, even when in this case there is no metric and therefore no notion 
of causality.} In the second one we consider physical observables and in the third part we construct the generators of gauge transformations.

\subsection{Canonical decomposition}

Since $\tilde{\varepsilon}^{abc}=3t^{[a}\tilde{\varepsilon}^{bc]}$ we have 
\begin{equation}
S_{\mr{CS}} = - \frac{1}{2}\int_{\mathcal{B}} \md^3 x\, \bigl( \tilde{\varepsilon}^{bc} \phi F_{bc}+
2\tilde{\varepsilon}^{ab} t^c F_{bc} A_a \bigr)
=  - \int_{\mathcal{B}}\md^3 x\,  \tilde{\varepsilon}^{ab} [(\pounds_{t} A_{a})A_b + F_{ab} \phi]\, ,
\end{equation}
where $\phi =t^aA_a$ and $t^c F_{bc}=-\pounds_{t} A_{b}+\nabla_b\phi$. The corresponding momenta are
\begin{equation}
\tilde{\Pi}^a=- \tilde{\varepsilon}^{ab}A_b\, ,
\end{equation}
and the canonical Hamiltonian is
\begin{equation}
H_{\mr{CS}}= \int_{\partial\Sigma} \md^{2}y\,  \left[ (\pounds_{t}\phi ) \tilde{\Pi}_{\phi} + ( \pounds_{t} A_{a} ) \tilde{\Pi}^{a} - \mathcal{L}_{CS}  \right]\approx
\int_{\partial\Sigma} \md^{2}y\, \tilde{\varepsilon}^{ab}F_{ab} \phi\, .
\end{equation}
Note that $H_{\mr{CS}}$ has the same form as the canonical Hamiltonian of the third approach on the bulk (\ref{CanonicalHamiltonian-PontryaginV3}),
but now it is differentiable for arbitrary variations $\delta A_a$ and $\delta\phi$, since there are no boundary terms (the boundary of the boundary is zero). 

The theory has three primary constraints
\begin{eqnarray}
\tilde{\Pi}_\phi & := & t_a\tilde{\Pi}^a\approx 0\, ,\label{CS.1stclass}\\
\varphi^a & := & \tilde{\Pi}^a +   \tilde{\varepsilon}^{ab}A_b\approx 0\, .\label{CS.2ndclass}
\end{eqnarray}
The constraints $\varphi^a$ are second class, since
\begin{equation}
\{  \varphi^a(y), \varphi^b(y')\} = 2 \tilde{\varepsilon}^{ab}\delta^2(y,y')\, .
\end{equation}
The total Hamiltonian is
\begin{equation}
H_{\mr{TCS}}= \int_{\partial\Sigma} \md^{2}y\, [\tilde{\varepsilon}^{ab}F_{ab} \phi +\lambda\tilde{\Pi}_\phi  + 
\lambda_a (\tilde{\Pi}^a +  \tilde{\varepsilon}^{ab}A_b)]\, , \label{TotalCS}
\end{equation}
and is differentiable for arbitrary fields variations and for arbitrary multipliers $\lambda$ and $\lambda_a$. The consistency conditions for
primary constraints lead to
\begin{eqnarray}
\{\tilde{\Pi}_\phi , H_{\mr{TCS}}\} & = &- \tilde{\varepsilon}^{ab}F_{ab}\approx 0\, ,\\ 
\{ \varphi^a, H_{\mr{TCS}}\} & = & 2 \tilde{\varepsilon}^{ab} (\partial_b\phi -\lambda_b)\, .
\end{eqnarray}
As a result, there is a secondary constraint
\begin{equation}
\Psi :=  \tilde{\varepsilon}^{ab}F_{ab}\approx 0\, ,\label{sec:constraint}
\end{equation}
and the multipliers $\lambda_a$ are determined as $\lambda_a = \partial_a\phi$. Now, the consistency condition for $\Psi$ is trivially fulfilled. If we consider the smeared constraint
\begin{equation}
\Psi [v] = \int_{\partial\Sigma} \md^{2}y\, v\,  \tilde{\varepsilon}^{ab}F_{ab}\approx 0\, ,
\end{equation}
its variation is given by
\begin{equation}
\delta  \Psi [v]=-2\int_{\partial\Sigma} \md^{2}y\, \tilde{\varepsilon}^{ab} \partial_av\, \delta A_b\, .
\end{equation}
Then,
\begin{equation}
\{\Psi [v],H_{\mr{TCS}}\} = 2\int_{\partial\Sigma} \md^{2}y\, \tilde{\varepsilon}^{ab}\partial_a\lambda_b = 2\int_{\partial\Sigma} \md^{2}y\, \tilde{\varepsilon}^{ab}\partial_a\partial_b\phi =0\, ,
\end{equation}
and there are no tertiary constraints. 

Let us now analyze the constrained structure of the theory. There are two first class constraints
\begin{eqnarray}
\tilde{\Pi}_\phi &\approx& 0\, ,\\
\mathcal{C} &:=& \nabla_a\varphi^a -\Psi=\nabla_a(\tilde{\Pi}^a-
\tilde{\varepsilon}^{ab}A_b ) \approx 0\, ,
\end{eqnarray}
and two second class constraints 
\begin{equation}
\varphi^a\approx 0\, .
\end{equation}
The counting of local degrees of freedom leads to  $3\cdot 2-2\cdot 2-2=0$. 

Note that, after introducing the expression for  $\lambda_a = \partial_a\phi$ and performing the integration by parts, the total Hamiltonian  (\ref{TotalCS}) can be rewritten as a combination of the first class constraints
\begin{equation}
H_{\mr{TCS}}= \int_{\partial\Sigma} \md^{2}y\, (- \phi\,  \mathcal{C}+\lambda\,\tilde{\Pi}_\phi )\, .
\end{equation}

We can obtain the equations of motion from the canonical action
\begin{equation}
 S_{\mr{CCS}}=\int_{\mathcal{B}}\md^3 x\,  [(\pounds_{t}\phi ) \tilde{\Pi}_{\phi} + (\pounds_{t} A_{a} ) \tilde{\Pi}^{a}
 +\phi\,  \mathcal{C} -\lambda\tilde{\Pi}_\phi ]\, .
\end{equation}
Then
\begin{eqnarray}
\frac{\delta S_{\mr{C}}}{\delta\tilde{\Pi}_{\phi}}=0\ \ &\Rightarrow& \ \ \pounds_{t}\phi =\lambda\, ,\\
\frac{\delta S_{\mr{C}}}{\delta\tilde{\Pi}^a}=0\ \ &\Rightarrow& \ \ \pounds_{t}A_a=\nabla_a\phi\, .
\end{eqnarray}
The last equation implies that $t^a F_{ab}=\pounds_{t}A_b-\nabla_b\phi=0$. Thus, we recover the Lagrangian equations of motion that state that the $3D$ connection $A_a$ on $\mathcal{B}$ is flat. The pullback of $F_{ab}$ to $\partial\Sigma$ is zero due to the constraint (\ref{sec:constraint}), and the transverse part vanishes as well as shown by the previous equation. Thus, the full $3D$ curvature vanishes $F_{ab}=0$, recovering the
covariant equations of motion for the Chern-Simons theory. 

Let us end this part with a remark. We have performed a complete Dirac analysis of the Abelian Chern-Simons theory on ${\mathcal{B}}$, sometimes referred as a `true-Dirac' analysis. 
This has to be compared to a reduced analysis that can sometimes be found in the literature, for instance in \cite{Blagojevic} and references therein. In this latter analysis, one solves  the
second class constraint (\ref{CS.2ndclass}) for $\tilde{\Pi}^a$, and gauge fixes the first class constraint (\ref{CS.1stclass}) in order to eliminate $\phi$ and its conjugate momentum as  dynamical
variables. The end result is that the (partially) reduced phase space consists of $U(1)$ connections
$A_a$, subject to the Dirac-Poisson bracket $\{ A_a(x),A_b(y)\}_{\mathrm{D}} =  \underaccent{\tilde}{\varepsilon}_{ab}\,\delta^2(x,y)$, 
and to the only remaining, first class constraint, $\Psi=\tilde{\varepsilon}^{ab}F_{ab}\approx 0$. The counting of local degrees of freedom, $2\cdot1 - 2\cdot1=0$ still yields zero. Needless to say, the final description of the physical theory is equivalent in both cases. We have chosen to perform the complete analysis in order to make contact with the variables and degrees of freedom of the bulk theory.

\subsection{Observables}

Just as we did in the previous section with the bulk theory, we shall attempt to construct observables
for the boundary theory. We know that there are no local degrees of freedom, so if there are non-trivial degrees of freedom they have to come from topological considerations. Since we are choosing a rather
trivial topology for the space ${\mathcal{B}}$, we do not expect to obtain non-trivial observables. Still it is a useful exercise to go through the process.

We start by constructing a linear functional
\begin{equation}
\mathcal{O}[f_a]=\int_{\partial\Sigma} \md^{2}y\,  f_a \tilde{\Pi}^a\, .
\end{equation}
Its Poisson bracket with the smeared first class constraint $\mathcal{C}$
\begin{equation}
C [g] =  \int_{\partial\Sigma} \md^{2}y\, g \nabla_a(\tilde{\Pi}^a-\theta
\tilde{\varepsilon}^{ab}A_b )\, ,
\end{equation}
is
\begin{equation}
\{ \mathcal{O}[f_a],C [g]\} =-\theta \int_{\partial\Sigma} \md^{2}y\, \tilde{\varepsilon}^{ab} f_a\nabla_b g\, ,
\end{equation}
and, since $g$ is an arbitrary function, $\mathcal{O}[f_a]$ is an observable for $f_a=\nabla_a f$, but it is a 
trivial one since $\nabla_a\tilde{\Pi}^a$ is a linear combination of constraints,
\begin{equation}
\mathcal{O}[f]=\int_{\partial\Sigma} \md^{2}y\,  (\nabla_a f) \tilde{\Pi}^a
=  -\int_{\partial\Sigma} \md^{2}y\, f \nabla_a\tilde{\Pi}^a
=  -\frac{1}{2} \int_{\partial\Sigma} \md^{2}y\, f (\mathcal{C}+\nabla_a\varphi^a)\approx 0         \, .
\end{equation}

In our search for observables,
one can also consider other combinations. Let us first note that $A_a\tilde{\Pi}^a\approx -\theta\tilde{\varepsilon}^{ab}A_aA_b=0$ and $F_{ab}\approx 0$.
Let us then construct another functional that also involves $\phi$ as follows,
\begin{equation}
\mathcal{P}[f]= \int_{\partial\Sigma} \md^{2}y\, f (\nabla_a\phi )\tilde{\Pi}^a\, .
\end{equation}
In order to be an observable its Poisson bracket with $C_1[g]= \int_{\partial\Sigma} \md^{2}y\,  g\tilde{\Pi}_{\phi}$ must vanish
\begin{equation}
\{  \mathcal{P}[f], C_1[g]\} =-\int_{\partial\Sigma} \md^{2}y\,  g\nabla_a (f\tilde{\Pi}^a)\approx 
-\int_{\partial\Sigma} \md^{2}y\,  g (\nabla_a f) \tilde{\Pi}^a\, ,
\end{equation}
and it vanishes only if $f={\mathrm{const}}$, but in that case
\begin{equation}
\mathcal{P}[f]= -f\int_{\partial\Sigma} \md^{2}y\, \phi (\nabla_a\tilde{\Pi}^a)\approx 0\, . 
\end{equation}
That is, we have found a trivial observable again. Thus, our limited search for physical observables takes us again to trivial observables. Note that since $\partial\Sigma$ does not have a boundary, we can not construct observables from the constraints by adding boundary terms.

\subsection{Generator of gauge transformations}

To end this section, let us find the generators of gauge transformation in the boundary theory.
The generators of gauge transformation are of the form
\begin{equation}
G[\epsilon_1,\epsilon_2]=\int_{\partial\Sigma} \md^{2}y\,  \left[ \epsilon_1\tilde{\Pi}_\phi + \epsilon_2 
\nabla_a(\tilde{\Pi}^a-\tilde{\varepsilon}^{ab}A_b)\right] \, ,
\end{equation}
where $\epsilon_1$ and $\epsilon_2$ are arbitrary (possibly phase space dependent) functions. The corresponding gauge transformations are
\begin{eqnarray}
\delta\phi &=& \epsilon_1\, ,\nonumber\\
\delta A_a &=& -\nabla_a \epsilon_2\, .
\end{eqnarray}
We can consider three special cases:
\begin{enumerate}
\item $U(1)$ gauge transformations can be obtained, for the choice $\epsilon_1 = \pounds_{t}\epsilon$ and $\epsilon_2=-\epsilon$, then
\begin{eqnarray}
\delta_{\epsilon}\phi &=& \pounds_{t}\epsilon\, ,\nonumber\\
\delta_{\epsilon} A_a &=& \nabla_a\epsilon\, .
\end{eqnarray}

\item Spatial diffeomorphisms. With the choice of field dependent parameter  $\epsilon_1 = \pounds_{\xi}\phi$ and $\epsilon_2=-\xi^a A_a$, such that $\xi^a t_a=0$, we obtain
\begin{eqnarray}
\delta_{\xi}\phi &=& \pounds_{\xi}\phi\, ,\nonumber\\
\delta_{\xi} A_a &=& \pounds_{\xi}A_a +\xi^b F_{ba}\approx \pounds_{\xi}A_a  \, ,
\end{eqnarray}
where the last equation is valid on the constrained phase space surface.

\item 'Time-like' diffeomorphisms. Time-like or time evolution diffeomorphisms can be obtained for $\epsilon_1 = \pounds_{t}\phi$ and $\epsilon_2=-\phi$. Thus,
\begin{eqnarray}
\delta_{t}\phi &=& \pounds_{t}\phi\, ,\nonumber\\
\delta_{t} A_a &=& \pounds_{t}A_a +t^b F_{ba}\approx \pounds_{t}A_a  \, ,
\end{eqnarray}
where the last equation is valid on-shell, that is, on the space of the solutions of the equations of motion. 
\end{enumerate}
Thus, we see that the physical theory is both ``gauge invariant" and diffeomorphism invariant on
$\mathcal{B}$, with no local degrees of freedom.

\section{Discussion on the canonical analysis: Pontryagin vs. Chern-Simons}
\label{sec:4}

In this section we would like to discuss and comment on the similarities and differences one encounters in analyzing both the bulk and the boundary theories. We summarize these results in the Table I below.

\begin{table}[h!]
\begin{center}
\begin{tabular}{|c|c|c|}
\hline
& \textbf{Pontryagin}&\textbf{Chern-Simons}\\
\hline
& & \\ 
\textbf{Action}&$S_{P} =  \frac{1}{4} \int_{\mathcal{M}} \md^{4} x\, \tilde{\varepsilon}^{abcd}F_{ab} F_{cd}$&
$S_{CS}=-\frac{1}{2}\int_{\mathcal{B}}\md^3 x\, \tilde{\varepsilon}^{abc} A_a F_{bc}$\\
& & \\\hline
& & \\
\textbf{Canonical variables}& 8-dim.\ \ $(\phi, \tilde{\Pi}_{\phi}; A_a,\tilde{\Pi}^{a})$  & 
6-dim. \ \ $(\phi, \tilde{\Pi}_{\phi}; A_a,\tilde{\Pi}^{a}$)\\ 
& & \\\hline
\textbf{Constraints}& & \\ 
Primary&$\tilde{\mathcal{C}}_1 := \tilde{\Pi}_{\phi} ,\ \ \ \tilde{\mathcal{C}}_2^a := \tilde{\Pi}^a - \tilde{\varepsilon}^{abc} F_{bc}$
&$\tilde{\mathcal{C}}_1 := \tilde{\Pi}_{\phi}, \ \ \ \varphi^a :=  \tilde{\Pi}^a +   \tilde{\varepsilon}^{ab}A_b$\\ 
Secondary&$\tilde{\mathcal{C}}_3:=\nabla_a\tilde{\Pi}^a$
&$\Psi := \tilde{\varepsilon}^{ab}F_{ab}$     \\ 
& & \\
1st class& $\tilde{\mathcal{C}}_1$,\ \ $\tilde{\mathcal{C}}_2^a$,\ \ $\tilde{\mathcal{C}}_3=\nabla_a\tilde{\mathcal{C}}_2^a$
& $\tilde{\mathcal{C}}_1$,\ \ $\mathcal{C} := \nabla_a\varphi^a -\Psi$          \\ 
2nd class& none &  $\varphi^a$  \\ 
& & \\\hline
 & & \\
\textbf{Total Hamiltonian}& $H_{\mr{T1}}= \int_{\Sigma} \md^{3}x\, (u\,\tilde{\mathcal{C}}_1      -\phi\,\tilde{\mathcal{C}}_3 
+ u_a\,  \tilde{\mathcal{C}}_2^a)$ & 
$H_{TCS}= \int_{\partial\Sigma} \md^{2}y\, (u\,\tilde{\mathcal{C}}_1 - \phi\, {\mathcal{C}})$\\
Diff. conditions& On $\partial\Sigma$: $F_{ab}= 0$;\  $u_a=0$ or 
$u_a=\partial_a f$       &none\\
& & \\\hline
& & \\ 
\textbf{Gauge generator}& $G=\int_{\Sigma}\md^3x\, [ \epsilon_1 \tilde{\mathcal{C}}_1+ 
\epsilon_2 \tilde{\mathcal{C}}_3+ \eta_a \tilde{\mathcal{C}}_2^a]$&
$G=\int_{\partial\Sigma} \md^{2}y\,  [ \epsilon_1 \tilde{\mathcal{C}}_1+ \epsilon_2\mathcal{C}]$ \\
Diff. conditions &On $\partial\Sigma$: $F_{ab}= 0$;\  $\eta_a=0$ or 
$\eta_a=\partial_a f$ & none\\
& & \\\hline
\textbf{Gauge symmetries}& & \\
U(1)& $\epsilon_1= \pounds_{t}\epsilon$, $\epsilon_2=-\epsilon$, $\eta_a=0$&
$\epsilon_1 = \pounds_{t}\epsilon$, $\epsilon_2=-\epsilon$   \\
Spatial Diffeo.& $\epsilon_1=\pounds_{\xi}\phi$, $\epsilon_2=-\xi^a A_a$, $\eta_a=\xi^b F_{ba}$
& $\epsilon_1 = \pounds_{\xi}\phi$, $\epsilon_2=-\xi^a A_a$\\
Time-like Diffeo.&  $\epsilon_1=\pounds_{t}\phi$, $\epsilon_2=-\phi$,
$\eta_a=\pounds_{t}A_a-\nabla_a\phi$ & $\epsilon_1 = \pounds_{t}\phi$, $\epsilon_2=-\phi$\\ 
& & \\\hline
\multirow{2}{*}{\textbf{Observables}}& & \\
&$\mathcal{O}_1[f]:=\int_{\Sigma}\md^3 x\, (\nabla_a f)\tilde{\Pi}^a\approx 0$
&$\tilde{\mathcal{O}}_1[f]:=
\int_{\partial\Sigma} \md^{2}y\,  (\nabla_a f )\tilde{\Pi}^a\approx 0$\\ 
&$\mathcal{O}_2:=\int_{\Sigma}\md^3 x\, (\nabla_a \phi)\tilde{\Pi}^a\approx 0$
&$\tilde{\mathcal{O}}_2:= \int_{\partial\Sigma} \md^{2}y\, (\nabla_a\phi )\tilde{\Pi}^a\approx 0$\\
& & \\\hline
\end{tabular}
\caption{Comparison between canonical descriptions of Pontryagin and Chern-Simons theories}
 
\end{center}
 
\end{table}

The Pontryagin term is defined on a 4-dim. manifold 
$\mathcal{M}$ with boundary $\partial\mathcal{M}=\mathcal{B}\cup\Sigma_1\cup\Sigma_2$, where 
$\mathcal{B}=I\times\partial\Sigma$ is a ``time-like" boundary, while the Chern-Simons term is defined on 
a 3-dimensional  $\mathcal{B}$. In the canonical Hamiltonian formulation of these theories the 
corresponding kinematical phase spaces are 8 and 6 dimensional, respectively. In the Pontryagin theory 
there  are four constraints, all of them are first class, while in the Chern-Simons theory there are two 
first class constraints and two second class ones. 

We found that the Hamiltonian of the Pontryagin theory is differentiable for arbitrary variations 
$\delta\phi$ only if $F_{ab}=0$ on the boundary $\partial\Sigma$. That condition appears also in the 
Chern-Simons theory, but in the form of a secondary constraint. Although we didn't a priori specify any 
boundary conditions, nor a condition on the variations that are allowed, the requirement of having a well 
defined Hamiltonian imposes a restriction on the variation of $A_a$; the only allowed variations are pure 
gauge.  

When analysing the gauge symmetries that both theories posses,
we see that they both have the same gauge symmetries, whose generators can be constructed out of the 
corresponding first class constraints. Again, even when the structure and the explicit form of the 
constraints is different in each case, we do recover in both cases, local $U(1)$ gauge transformations, 
spatial diffeomorphisms, and transverse (``time-like") diffeomorphisms. Thus, in both cases we
recover a theory of gauge invariant, flat, and diffeomorphism invariant connections on $\partial\Sigma$.

Finally, we constructed and analysed physical observables for both theories, and found in both cases  the 
same type of observables, even though all of them are trivial, due to the triviality of the topologies 
chosen.  Here we have taken $\Sigma$ such that $\partial\Sigma=S^2$. Had we chosen a non-trivial boundary, 
we would have non-trivial global degrees of freedom (accessible through Wilson-loops around homotopically 
nontrivial loops, for instance).  As we discussed in the previous section, the observables in both 
theories can be written as integrals over the boundary, in such a way that can easily be related and 
identified with each other (note that in the table the Pontryagin observables have been rewritten as bulk 
integrals).

Let us end this with a remark. As we mentioned in the introduction, in the manuscript \cite{EC} these two 
theories have been studied in the canonical formalism (for $SO(3,1)$), which makes their results very 
close to ours. However, there are important and subtle differences that we enumerate as follows:
(1) Here the Pontryagin theory is actually defined on a manifold with a boundary and the Chern-Simons 
theory is defined on this boundary; (2) We rewrite the Hamiltonian of the Pontryagin theory in three 
equivalent ways, inspect the resulting boundary conditions  and relate them to constraints of the Chern-
Simons theory; (3) We use a fully covariant approach in the canonical decomposition, without fixing a 
foliation nor a coordinate system and; (4) We construct observables in both theories that allow us to make 
direct comparison between the two theories. 

\section{Outlook}
\label{sec:5}

Gauge theories defined on regions with boundaries are without doubt, of wide physical interest. A complete 
canonical description of those theories based on the Dirac algorithm is, in our opinion, still lacking 
(see however \cite{MPB,Troessaert}). This is particularly noticeable when one has a theory defined by an 
action with both bulk and boundary contributions, and this later possesses time derivatives of the 
fundamental variables. In this case, the corresponding canonical momenta have contributions from both the 
bulk and the boundary, and one has then to properly define the symplectic and Poisson structures of the 
theory.  Our ultimate goal is to provide a systematic analysis of such scenarios.

Here we have considered a simpler case, where the theory can be alternatively seen as a bulk theory or as one defined on the boundary. What we have here learned will be useful in the next step that is coupling the Pontryagin bulk theory to a Maxwell field (or in the non-Abelian case, Yang Mills), and thus obtaining
the well studied case of Yang-Mills with a $\theta$ term.  
Here the challenge is to alternatively describe the resulting theory as a pure bulk theory or as having a 
bulk and a boundary contribution. Of course, both descriptions should coincide and one should then have an 
appropriate ``dictionary" (to borrow a term from AdS/CFT) to compare both theories. The relation between 
the two descriptions in our simple example should then be helpful to construct such dictionary for the 
Maxwell-Pontryagin case.

Of course, the system we have here analysed is rather simple, and in certain sense, trivial. Still, this 
feature that allows one to ``solve it", is relevant since one can then precisely point out how to relate 
and compare the two theories defined on different manifolds, bulk and boundary, and show their 
equivalence. We have also pointed out certain subtleties that have allowed us to make this comparison, 
such as a careful analysis of the boundary conditions imposed in order to make the canonical description of the 
bulk theory well defined, and how those boundary conditions are related to (secondary) constraints on the 
boundary. While such relations had been
suggested before in other systems, we feel that the simplicity of this one allows for a very clear 
understanding of this correspondence. 

We hope that these results will be useful in our program of understanding, from the ``Dirac" perspective,
gauge systems defined on regions with boundaries. This viewpoint has to be contrasted to, say, the geometrical approach of \cite{Barbero}, that can be seen as complementary to ours. We shall report some of those results elsewhere \cite{CV-MYM}.

\begin{acknowledgments}
We would like to thank I. Rubalcava-Garcia for discussions.
This work was in part supported by  CONACyT 0177840 and PAPIIT IN100218 grants 
and by CIC, UMSNH.
\end{acknowledgments}

\end{document}